\documentclass[a4paper,conference]{IEEEtran}
\usepackage{amsmath}%
\usepackage{mathtools} 
\usepackage{amsfonts}%
\usepackage{dsfont}%
\usepackage{amssymb}%
\usepackage{graphicx}
\usepackage{cite}
\usepackage{verbatim}%
\usepackage{enumerate}%
\usepackage[hmargin=0.65in , vmargin=1in]{geometry}
\usepackage{amsthm}
\usepackage{algorithm}
\usepackage{algorithmic}

\newtheorem{thm}{Theorem}

\newenvironment{prf}{\par{\noindent \it Proof:}}{\qed\par}
\newenvironment{proofsketch}{\par{\noindent \it Proof Sketch:}}{\qed\par}
\ifCLASSINFOpdf
\else

\fi

\title{Throughput Maximization in Multi-Channel Cognitive Radio Systems with Delay Constraints}

\author{Ahmed Ewaisha and Cihan Tepedelenlio\u{g}lu\\
\small{School of Electrical, Computer, and Energy Engineering, Arizona State University, USA.}\\
\small{Email:\{ewaisha, cihan\}@asu.edu}}

\begin{document}

\maketitle

\begin{abstract}
We study the throughput-vs-delay trade-off in an overlay multi-channel single-secondary-user cognitive radio system. Due to the limited sensing capabilities of the cognitive radio user, channels are sensed sequentially. Maximizing the throughput in such a problem is well-studied in the literature. Yet, in real-time applications, hard delay constraints need to be considered besides throughput. In this paper, optimal stopping rule and optimal power allocation are discussed to maximize the secondary user's throughput, subject to an average delay constraint. We provide a low complexity approach to the optimal solution of this problem. Simulation results show that this solution allows the secondary user to meet the delay constraint without sacrificing throughput significantly. It also shows the benefits of the optimal power allocation strategy over the constant power allocation strategy.
\end{abstract}

\section{Introduction}
\label{Introduction}
Cognitive Radio (CR) systems have been proposed to help overcome spectral inefficiency by allowing unlicensed users, called the Secondary Users (SU), to access the spectrum with minimal or with no degradation of the licensed users' performance. To guarantee this, the SU radios are equipped with sensing units capable of detecting the presence of the Primary Users' (PU) transmission activities over the spectrum. when no activity is detected at some frequency band, the channel is deemed free and the SU is allowed to use it.

If the PU can tolerate some interference, the SU will be allowed to transmit on busy frequency channels as long as the interference does not exceed this tolerance value. This scenario has been studied in \cite{Adaptive_Rate_Power_CR_Sonia} for a single channel case. The authors found an optimal closed-form expression to the amount of power that the SU can transmit under this interference constraint. In a multi-channel system, the problem becomes more interesting since the SU faces heterogeneous interference constraints. In this case, the SU will have to decide how much power should be allocated to each channel. This problem has been studied in \cite{Power_Knapsack_OFDM_CR} with the objective of maximizing the SU's throughput under per-channel interference constraints.

In some scenarios the SU may have limited sensing capabilities so that only one channel can be sensed at a time. In \cite{POMDP_Qing_Zhao} the authors considered this problem and showed how to optimally select this channel to maximize the SU's throughput. In a time-slotted system, if a channel is detected busy, then the time-slot is wasted. Thus, in \cite{Sensing_Order_Poor} the authors allowed multiple channels to be sensed in the same time-slot sequentially. Because the duration of sensing a channel may be large, the SU is allowed to stop sensing at some channel and begin transmission. The authors discussed the optimal stopping rule and proposed a dynamic programming algorithm to find the optimal sensing order that maximizes the SU's throughput. The channels were assumed to have heterogeneous channel availability probabilities. Their work was extended in \cite{Ewaisha_First} to optimize over the channel sensing duration that was assumed fixed in \cite{Sensing_Order_Poor}. Optimal Stopping Rule has been considered in \cite{Polynomial_Alg_Optimal_Stopping_Rule} as well, where the authors generalized the setup to heterogeneous fading channels and devised a polynomial-time algorithm to find the optimal stopping rule.

In real-time applications, such as video and audio streaming, delay is an important factor in evaluating the performance of a communication system, because packets are expected to arrive at the destination before some delay constraint. In this work we study the fundamental trade-off between throughput and delay in a CR context. We propose a channel sensing and access scheme and find the optimal stopping rule that maximizes a single SU's throughput while maintaining delay bounds on the packets of the CR user. To the best of our knowledge, throughput-delay trade-off has not been considered before in multichannel CR systems.

The rest of this paper is organized as follows. We present our system model in Section \ref{System_Model}. In Section \ref{Problem_Formulation} we formulate the problem mathematically. The optimal solution is discussed in Section \ref{Optimal_Solution} by solving a special case of the problem to get some insights, then we present the optimal solution to the original problem along with a discussion of our low-complexity approach. Simulation results are presented in Section \ref{Simulation_Results}, and the paper is concluded in Section \ref{Conclusion}.

\section{System Model}
\label{System_Model}
We adopt a time-slotted system with slot duration of $T$ seconds. The PU has $M$ licensed frequency channels to access. We assume that the PU's transmission begins only at the beginning of a time-slot, and stops at the end of this time-slot with the opportunity to transmit on subsequent time-slots independently. PU's transmission is independent across channels and across time-slots. On the other hand, we have a single SU trying to access one of these $M$ channels. Before a time-slot begins, the SU is assumed to have ordered the channels according to some sequence\footnote{The method of ordering the channels is outside the scope of this work. The reader is referred to \cite{Sensing_Order_Poor} for further details about channel ordering.}. We denote the first channel in this sequence as channel $1$, the second channel as $2$, and so on until the last channel being $M$. Before the SU attempts to transmit its packet over channel $i$, it senses this channel to determine its ``state'' which is described by a Bernoulli random variable $b_i$ with parameter $\theta_i$. If $b_i=1$ (which happens with probability $\theta_i$), then channel $i$ is free and the SU can transmit over it until the on-going time-slot ends. If $b_i=0$, channel $i$ is busy, and the SU proceeds to sense channel $i+1$.

We assume that the SU's sensing errors are negligible but the SU has limited capabilities in the sense that no two channels can be sensed simultaneously. This may be the case when considering radios having a single sensing module with a fixed bandwidth, so that it can be tuned to only one frequency channel at a time. Therefore, at the beginning of a given time-slot, the SU selects a channel, say channel $1$, senses it for $\tau$ seconds ($\tau \ll T/M$), and transmits over this channel if it is free. Otherwise, the SU skips this channel and senses channel $2$, and so on until it finds a free channel. If all channels are busy (i.e. the PU has transmission activities on all $M$ channels) then this time-slot will be considered as ``wasted''. In this case, the SU waits for the following time-slot and begins sensing following the same channel sensing sequence.

\begin{figure}
	\centering
		\includegraphics[width=0.9\columnwidth]{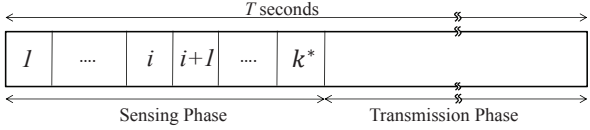}
	\caption{Sensing and transmission phases in one time-slot. In this example, transmission begins after sensing the $k^*$th frequency channel.}
	\label{Time_Slot_Fig}
\end{figure}

The physical channel between the SU transmitter and corresponding receiver is assumed to be flat fading with independent, identically distributed (i.i.d.) channel gains among the $M$ channels. To achieve higher data rates, the SU uses adaptive modulation. This means that the SU adapts the data rate according to the instantaneous power gain of the channel before beginning transmission on this channel. To do this, once the SU finds a free channel, say channel $i$, the gain $\gamma_i$ is probed. The data rate will be proportional to $\log(1+P_i(\gamma_i){\gamma}_i)$, where $P_i(\gamma_i)$ (or simply $P_i$) is the power transmitted by the SU at channel $i$ as a function of the instantaneous gain. Fig. \ref{Time_Slot_Fig} shows a potential scenario where the SU senses $k^*$ channels, skips the first $k^*-1$, and uses the $k^*$th channel for transmission until the end of this on-going time-slot. In this scenario the SU ``stops'' at the $k^*$th channel. Clearly, $k^* \in \{1,...,M\}$ is a random variable that changes from one time-slot to another. It depends on the states $[b_1,...,b_M]$ along with the gains of each channel $[\gamma_1,...,\gamma_M]$. To understand why, consider that the SU senses channel $i$, finds it free and probes its gain $\gamma_i$. If ${\gamma}_i$ is found to be low, then the SU skips channel $i$ (although free) and senses channel $i+1$. This is to take advantage of the possibility that ${\gamma}_{i+1}\gg{\gamma}_i$. On the other hand, if $\gamma_i$ is sufficiently large, the SU stops at channel $i$ and begins transmission. In that latter case $k^*=i$. Defining the two random vectors $\underline{b}=[b_1,...,b_M]^T$ and $\underline{\gamma}=[\gamma_1,...,\gamma_M]^T$, $k^*$ is a deterministic function of $\underline{b}$ and $\underline{\gamma}$.\\
We define the stopping rule by defining a threshold $\gamma_{\rm th}(i)$ to which each $\gamma_i$ is compared when the $i$th channel is found free. If $\gamma_i>\gamma_{\rm th}(i)$, the SU stops and transmits at channel $i$. Otherwise, channel $i$ is skipped and channel $i+1$ sensed. If $\gamma_{\rm th}(i)=0$, then whenever channel $i$ is sensed free, the SU will not skip it. Increasing $\gamma_{\rm th}(i)$ allows the SU to skip channel $i$ whenever ${\gamma}_i<\gamma_{\rm th}(i)$, to search for a better channel, thus potentially increasing the throughput. Setting $\gamma_{\rm th}(i)$ too large allows channel $i$ to be skipped even if $\gamma_i$ is high. This constitutes the trade-off on choosing the values of $\gamma_{\rm th}(i)$'s. The optimal values of $\gamma_{\rm th}(i)$ $i=1,...,M$, determine the optimal stopping rule.

\subsection{Average Power}
Now we will formulate the problem under some average power constraint. Let $P_i$ denote the power used at the $i$th channel, which is zero unless $i=k^*$.
Defining $c_{i} \coloneqq 1-\frac{i\tau}{T}$ which is the fraction of the time-slot remaining for the SU's transmission if the SU transmits on the $i$th channel in the sensing sequence. The average power constraint is $\mathbb{E} [c_{k^*} P_{k^*}(\gamma_{k^*})] \leq P_{\rm avg}$, where the expectation is over $\underline{b}$ and $\underline{\gamma}$. This expectation can be calculated recursively. Let
\begin{equation}
\begin{array}{ll}
S_i({\bf \Gamma}^{\rm th}_i,{\bf P}_i)&=\theta_i c_i \int_{\gamma_{\rm th}(i)}^\infty{P_i f_{\gamma_i}(\gamma) \,d\gamma}+
\\ & \left[1-\theta_i \bar{F}_\gamma(\gamma_{\rm th}(i)) \right]S_{i+1}({\bf \Gamma}^{\rm th}_{i+1},{\bf P}_{i+1}),
\label{Average_Power}
\end{array}
\end{equation}
$i=1...,M$, with $S_{M+1}({\bf \Gamma}^{\rm th}_{M+1},{\bf P}_{M+1})  \coloneqq  0$, while ${\bf P}_i \coloneqq [P_i,...,P_M]^T$, ${\bf \Gamma}^{\rm th}_i \coloneqq [\gamma_{\rm th}(i),...,\gamma_{\rm th}(M)]^T$, and $f_{\gamma_i}(\gamma)$ is the Probability Density Function (PDF) of the gain $\gamma_i$ of channel $i$, and $\bar{F}_{\gamma_i}(x) \coloneqq \int_x^{\infty}{f_{\gamma_i}(\gamma) \, d\gamma}$. The first term in (\ref{Average_Power}) is the average power transmitted at channel $i$ given that channel is chosen for transmission (i.e. given that $k^*=i$). The second term represents the case where channel $i$ is skipped and channel $i+1$ is sensed. It can be shown that
\begin{equation}
S_1({\bf \Gamma}^{\rm th}_1,{\bf P}_1)=\mathbb{E} \left[c_{k^*} P_{k^*} \right].
\label{Avg_Pow_Recursive_Formula}
\end{equation}
We note that we will drop the index $i$ from the subscript of $f_{\gamma_i}(\gamma)$ and $\bar{F}_{\gamma_i}(\gamma)$ since channels suffer i.i.d. fading.

\subsection{Throughput}
The SU's average throughput is defined as $\mathbb{E} [c_{k^*} \log(1+P_{k^*}\gamma_{k^*})]$. Similarly, we denote the expected throughput as $U_1({\bf \Gamma}_1^{th},{\bf P}_1)$ which can be derived using the following recursive formula
\begin{equation}
\begin{array}{ll}
U_i({\bf \Gamma}^{\rm th}_{i},{\bf P}_i)&=\theta_i c_i \int_{\gamma_{\rm th}(i)}^{\infty}{\log(1+P_i\gamma) f_{\gamma}(\gamma)} \, d\gamma
\\ & + \theta_i U_{i+1}({\bf \Gamma}^{\rm th}_{i+1},{\bf P}_{i+1}) \int_0^{\gamma_{\rm th}(i)}{f_{\gamma}(\gamma)} \, d\gamma
\\ & + (1-\theta_i)U_{i+1}({\bf \Gamma}^{\rm th}_{i+1},{\bf P}_{i+1}),
\label{Reward}
\end{array}
\end{equation}
$i=1,...,M$, with $U_{M+1} \coloneqq 0$.

\subsection{Delay}
If the SU skips all channels, either due to being busy or due to their low gain, then this time-slot is wasted and a \emph{blocking} event occurs. The SU has to wait for the following time-slot to begin searching for a free channel again. This results in a delay to the SU's transmitted packet. In real-time applications, there may exist some average delay requirement $\bar{D}_{\rm{max}}$ on each packet. That is, the average packet delay may not exceed $\bar{D}_{\rm{max}}$.

Define the delay $D$ as the number of time-slots the SU wastes, due to blocking, while trying to transmit a given packet. Here, we put a constraint on the expected value of $D$, $\mathbb{E}[D]$. Since the availability of each channel is independent across time-slots, $D$ follows a geometric distribution having $\mathbb{E}[D]= \left(\rm{Pr}[\rm{Success}]\right)^{-1}$ where $\rm{Pr}[\rm{Success}]=1-\rm{Pr}[\rm{Blocking}]$ is the probability of not wasting a time-slot, i.e. no blocking occurs. In other words, $\rm{Pr}[\rm{Success}]$ is the probability that the SU finds a free channel having a high gain that makes it not skipped, out of the $M$ possible channels. It is given by $\rm{Pr}[\rm{Success}]  \coloneqq  p_1({\bf \Gamma}^{\rm th}_1)$ which is calculated recursively using the following equation
\begin{align}
\nonumber p_i({\bf \Gamma}^{\rm th}_i)&=\theta_i \bar{F}_\gamma(\gamma_{\rm th}(i))+ \left[1-\theta_i \bar{F}_\gamma(\gamma_{\rm th}(i)) \right]p_{i+1}({\bf \Gamma}^{\rm th}_{i+1}), 
\label{Prob_recursive}
\end{align}
$i=1,...,M$, where $p_{M+1} \coloneqq 0$. Here, $p_i({\bf \Gamma}^{\rm th}_i)$ is the probability that the SU transmits on channel $i$, $i+1$,..., or $M$.

\section{Problem Statement}
\label{Problem_Formulation}
From equation (\ref{Reward}) we can see that the SU's expected throughput, $U_1$, is affected by the thresholds $\gamma_{\rm th}(i)$'s. The goal is to find the optimum value of $\gamma_{\rm th}(i)$'s that maximizes $U_1$ subject to an expected packet delay constraint. The delay constraint can be written as $\mathbb{E}[D] \leq \bar{D}_{\rm{max}}$ or, equivalently, $p_1({\bf \Gamma}^{\rm th}_1) \geq \frac{1}{\bar{D}_{\rm{max}}}$. Mathematically, the problem becomes

\begin{equation}
\begin{array}{ll}
\rm{maximize}& U_1({\bf \Gamma}^{\rm th}_1,{\bf P}_1)\\
\label{Optimization_Problem_Optimum_Pow_Control}
\rm{subject \; to} &S_1({\bf \Gamma}^{\rm th}_1,{\bf P}_1) \leq P_{\rm{avg}}\\
& p_1({\bf \Gamma}^{\rm th}_1) \geq \frac{1}{\bar{D}_{\rm{max}}}\\
\rm{variables} & {\bf \Gamma}^{\rm th}_1,{\bf P}_1,
\end{array}
\end{equation}
where the first constraint represents the average power constraint, while the second is a bound on the average packet delay.

\section{Optimal Solution}
\label{Optimal_Solution}
\subsection{Two-Level Power Control: A Special Case}
First we solve this problem assuming $P_i=1$ for $i=k^*$ and $P_i=0$ for $i \neq k^*$ as a special case to get some insights, then we generalize the solution to the more general case where $P_i \equiv P_i(\gamma_i)$ are functions of the channel gain. The optimization problem in (\ref{Optimization_Problem_Optimum_Pow_Control}) becomes
\begin{equation}
\begin{array}{ll}
\rm{maximize}& U_1({\bf \Gamma}^{\rm th}_1,{\bf 1}_M)\\
\label{Optimization_Problem_2_Level_Pow_Control}
\rm{subject \; to} & p_1({\bf \Gamma}^{\rm th}_1) \geq \frac{1}{\bar{D}_{\rm{max}}}\\
\rm{variables} & {\bf \Gamma}^{\rm th}_1,
\end{array}
\end{equation}
where ${\bf 1}_M$ is a vector with $M$ ones. Let $\lambda_{\rm D}$ be the Lagrange multiplier associated with the delay constraint in (\ref{Optimization_Problem_2_Level_Pow_Control}). The Lagrangian associated with problem (\ref{Optimization_Problem_2_Level_Pow_Control}) is 
\begin{equation}
L({\bf \Gamma}^{\rm th}_1,\lambda_{\rm D})=U_1\left({\bf \Gamma}^{\rm th}_1,{\bf 1}_M \right)+\lambda_{\rm D} \left( p_1({\bf \Gamma}^{\rm th}_1) - \frac{1}{\bar{D}_{\rm{max}}} \right).
\label{Lagrange_2_Level_Pow_Control}
\end{equation}
To solve problem (\ref{Optimization_Problem_2_Level_Pow_Control}) we find the KKT equations that represent necessary conditions for the optimal solution. The first equation among those KKT equations is obtained by differentiating (\ref{Lagrange_2_Level_Pow_Control}) with respect to $\gamma_{\rm th}(i)$ for each $i \in \{1,...,M\}$ and equating each derivative to zero. Thus we get
\begin{align}
\nonumber & \gamma_{\rm th}^*(i)=\\
&\left[\exp\left(\frac{U_{i+1}({\bf \Gamma}^{\rm th*}_{i+1},{\bf 1}_{M-i})-\lambda_{\rm D}^* \cdot \left(1-p_{i+1}^* \right)}{c_i} \right)-1 \right]^+,
\label{Gamma_Solution}
\end{align}
$i=1,...,M$, where $(x)^+ \coloneqq  {\rm{max}}(x,0)$, ${\bf \Gamma}^{\rm th*}_{i+1}  \coloneqq  [\gamma_{\rm th}^*(i+1),...,\gamma_{\rm th}^*(M)]^T$ while $p_{i+1}^* \coloneqq p_{i+1}\left({\bf \Gamma}^{\rm th *}_{i+1}\right)$, $p_{M+1}^* \coloneqq 0$ and $U_{M+1}({\bf \Gamma}^{\rm th*}_{M+1},{\bf 1}_0) \coloneqq 0$. The second KKT equation is $\lambda_{\rm D}^* \cdot \left(p_1^*-\frac{1}{\bar{D}_{\rm max}} \right)=0$. Thus, $\lambda_D^*$ is chosen to satisfy the inequality constraint in (\ref{Optimization_Problem_2_Level_Pow_Control}) with equality (i.e. $p_1^*=\frac{1}{\bar{D}_{\rm max}}$). Hence, we substitute by $\gamma_{\rm th}^*(i)$ in $p_1({\bf \Gamma}^{\rm th}_1)$ to get $p_1^*$ as a function of $\lambda_{\rm D}^*$. Then we solve for $\lambda_{\rm D}^*$ to satisfy $p_1^*=\frac{1}{\bar{D}_{\rm max}}$. The optimality of our proposed approach is discussed in the following theorem.
\begin{thm}
\label{Thm_Optimality_of_2_Level_Pow_Control}
The expression for $\gamma_{\rm th}^*(i)$ in (\ref{Gamma_Solution}), along with $\lambda_{\rm D}^*$ that solves $p_1\left({\bf \Gamma}_1^{\rm th} \right)=\bar{D}_{\rm max}^{-1}$, represent the optimal solution to problem (\ref{Optimization_Problem_2_Level_Pow_Control}).
\end{thm}

\begin{proofsketch}
The proof depends on the fact that $p_1^*$ is strictly monotonic in $\lambda_{\rm D}^*$. This indicates that there exists a unique $\lambda_{\rm D}^*$ satisfying $p_1^*=\frac{1}{\bar{D}_{\rm max}}$. Thus $\gamma_{\rm th}^*(i)$ is unique. Consequently, the KKT equations are sufficient for optimality.
\end{proofsketch}

\subsection{Optimal Power Control: General Case}
Now we allow the power $P_i$ to be an arbitrary function of $\gamma_i$ and optimize over this function to maximize the throughput subject to average power and delay constraints. The optimization problem is expressed by (\ref{Optimization_Problem_Optimum_Pow_Control}). Let $\lambda_{\rm P}$ and $\lambda_{\rm D}$ be the dual variables associated with the constraints in problem (\ref{Optimization_Problem_Optimum_Pow_Control}). The Lagrangian for (\ref{Optimization_Problem_Optimum_Pow_Control}) becomes
\begin{equation}
\begin{array}{lll}
L\left({\bf \Gamma}_1^{\rm th},{\bf P}_1, \lambda_{\rm P}, \lambda_{\rm D} \right)&=U_1 \left( {\bf \Gamma}^{\rm th}_1,{\bf P}_1 \right)-\\
&\lambda_{\rm P} \left( S_1({\bf \Gamma}^{\rm th}_1,{\bf P}_1) -P_{\rm{avg}}\right)+ \\
&\lambda_{\rm D} \left( p_1({\bf \Gamma}^{\rm th}_1) - \frac{1}{\bar{D}_{\rm{max}}} \right).
\end{array}
\label{Lagrange_Optimum_Pow_Control}
\end{equation}
The KKT conditions \cite{Cvx_Boyd} are necessary equations for the optimal point of problem (\ref{Optimization_Problem_Optimum_Pow_Control}). To get the KKT equations associated with this problem, we differentiate (\ref{Lagrange_Optimum_Pow_Control}) with respect to (w.r.t.) each of the primal variables $P_i$ and $\gamma_{\rm th}(i)$ and equate the resulting derivatives each to zero. We also have the primal feasible, dual feasible and complementary slackness equations. Thus the KKT equations are
\begin{align}
	\label{Water_Filling}
	&P_i^*(\gamma)=\left(\frac{1}{\lambda_{\rm P}^*} - \frac{1}{\gamma}\right)^+, \hspace{0.5cm} i=1,...,M, \\
	\label{gamma_i_Equation}
	\nonumber	&c_i \left( \log \left(\frac{\gamma_{\rm th}^*(i)}{\lambda_{\rm P}^*} \right) - \lambda_{\rm P}^* \left(\frac{1}{\lambda_{\rm P}^*} - \frac{1}{\gamma_{\rm th}^*(i)} \right)^+ \right)= \\
	&U_{i+1}^* - \lambda_{\rm P}^* S_{i+1}^*-\lambda_{\rm D}^* \cdot \left( 1-p_{i+1}^* \right), \hspace{0.25cm} i=1,...,M,\\
%
%
	&S_1^* \leq P_{\rm{avg}} \hspace{0.2cm}, \hspace{0.2cm} p_1^* \geq \frac{1}{\bar{D}_{\rm{max}}}\\
	& \lambda_{\rm P}^* \geq 0 \hspace{0.2cm}, \hspace{0.2cm} \lambda_{\rm D}^* \geq 0 \\
	\label{Comp_Slackness_Power}
	& \lambda_{\rm P}^* \cdot \left( S_1^* -P_{\rm{avg}}\right)=0\\
	\label{Comp_Slackness_Delay}
	&\lambda_{\rm D}^* \cdot \left( p_1^* - \frac{1}{\bar{D}_{\rm{max}}} \right)=0,
\end{align}
%
where ${\bf P}_{i+1}^*  \coloneqq  [P_{i+1}^*,...,P_M^*]^T$ is the optimal power allocation vector. We use $U_{i+1}^* \coloneqq U_{i+1}\left({\bf \Gamma}_{i+1}^{\rm th *},{\bf P}_{i+1}^* \right)$, $U_{M+1}^* \coloneqq 0$, $S_{i+1}^* \coloneqq S_{i+1}\left({\bf \Gamma}_{i+1}^{\rm th *},{\bf P}_{i+1}^* \right)$, $S_{M+1}^* \coloneqq 0$ and $p_{M+1}^* \coloneqq 0$ in the sequel.

\subsubsection{Primal Variables}
\label{Primal_Variables}
We can see that equation (\ref{Water_Filling}) is the \emph{water-filling} strategy that gives the closed-form optimum solution for the first $M$ primal variables, namely $P_i$ ($i=1,...,M$).
The remaining $M$ primal variables, namely $\gamma_{\rm th}^*(i)$ ($i=1,...,M$), are found via the set of equations (\ref{gamma_i_Equation}). Note that equations (\ref{gamma_i_Equation}) indeed form a set of $M$ equations, each solves for one of the $\gamma_{\rm th}^*(i)$'s ($i=1,...,M$). We refer to this set as the ``$\gamma$-finding'' equations. We also note that for a given value of $i$, solving for $\gamma_{\rm th}^*(i)$ needs the knowledge of only $\gamma_{\rm th}^*(i+1)$ through $\gamma_{\rm th}^*(M)$, and does not require knowing $\gamma_{\rm th}^*(1)$ through $\gamma_{\rm th}^*(i-1)$. In other words, the equation that finds $\gamma_{\rm th}^*(i)$ is a function of only $\gamma_{\rm th}^*(i)$ through $\gamma_{\rm th}^*(M)$. Thus, these $M$ equations can be solved using back-substitution. The optimal solution for the $\gamma$-finding equations is given by
\begin{align}
\nonumber & \gamma_{\rm th}^*(i)=\\
&\frac{-\lambda_{\rm P}^*}{W_0 \left[ -\exp \left(-\frac{\left[U_{i+1}^* - \lambda_{\rm P}^* S_{i+1}^*-\lambda_{\rm D}^* \left( 1-p_{i+1}^* \right)\right]^+}{c_i}-1\right) \right]},
\label{Gamma_Solution_Lambert_W}
\end{align}
$i=1,...,M$, and $W_0[z]$ is called the principle value of the Lambert W function \cite{Lambert_W_Function} and is given by
\begin{equation}
W_0[z]=\sum_{n=1}^{\infty} \frac{\left( -n \right)^{n-1}}{n!}z^n.
\label{Lambert_W_0}
\end{equation}
Although the $\gamma$-finding equations are necessary equations for the optimal point, Theorem \ref{Thm_Unique_Solution} proves that the solution ${\bf \Gamma}_1^{\rm th *}  \coloneqq  [\gamma_{\rm th}^*(1),...,\gamma_{\rm th}^*(M)]^T$ is unique given the dual variables $(\lambda_{\rm P},\lambda_{\rm D})$. This indicates that these equations are sufficient as well, proving that the solution to the $\gamma$-finding equations is optimum for problem (\ref{Optimization_Problem_Optimum_Pow_Control}) given the optimum dual variables $(\lambda_{\rm P}^*,\lambda_{\rm D}^*)$.
\begin{thm}
\label{Thm_Unique_Solution}
Given some value for $(\lambda_{\rm P}^*,\lambda_{\rm D}^*)$, the left-hand-side (l.h.s.) of equation (\ref{gamma_i_Equation}) is monotonically increasing in $\gamma_{\rm th}^*(i)$. Thus the solution of $\gamma_{\rm th}^*(i)$ that satisfies equation (\ref{gamma_i_Equation}) is unique.
\end{thm}
\begin{prf}
Let the l.h.s. of equation (\ref{gamma_i_Equation}) be
\begin{equation}
Y \left(\gamma_{\rm th}^*(i) \right)=c_i \left( \log \left(\frac{\gamma_{\rm th}^*(i)}{\lambda_{\rm P}^*} \right) - \lambda_{\rm P}^* \left(\frac{1}{\lambda_{\rm P}^*} - \frac{1}{\gamma_{\rm th}^*(i)}\right)^+ \right).
\label{Left_Hand_Side}
\end{equation}
If $\gamma_{\rm th}^*(i) < \lambda_{\rm P}^*$. Then $Y \left (\gamma_{\rm th}^*(i)\right)=c_i \log \left(\frac{\gamma_{\rm th}^*(i)}{\lambda_{\rm P}^*} \right)$ which is an increasing function in $\gamma_{\rm th}^*(i)$. On the other hand, if $\gamma_{\rm th}^*(i) > \lambda_{\rm P}^*$, $Y \left (\gamma_{\rm th}^*(i)\right)$ becomes
\begin{equation}
Y \left(\gamma_{\rm th}^*(i) \right)=c_i \left( \log \left(\frac{\gamma_{\rm th}^*(i)}{\lambda_{\rm P}^*} \right) -  \left(1 - \frac{\lambda_{\rm P}^*}{\gamma_{\rm th}^*(i)}\right) \right).
\end{equation}
Differentiating $Y(\gamma_{\rm th}^*(i))$ w.r.t. $\gamma_{\rm th}^*(i)$ we get
\begin{align}
\frac{\partial Y}{\partial \gamma_{\rm th}^*(i)}&=c_i \left(\frac{1}{\gamma_{\rm th}^*(i)} - \frac{\lambda_{\rm P}^*}{\gamma_{\rm th}^2(i)}\right)\\
& > c_i \left(\frac{1}{\gamma_{\rm th}^*(i)} - \frac{\lambda_{\rm P}^*}{\lambda_{\rm P}^*\gamma_{\rm th}^*(i)}\right)\\
& = 0.
\label{Derivative_of_Y}
\end{align}
This means that $Y(\gamma_{\rm th}^*(i))$ is monotonic in $\gamma_{\rm th}^*(i)$. Hence, $\gamma_{\rm th}^*(i)$ satisfying equation (\ref{gamma_i_Equation}) is unique.
\end{prf}
To solve any of the $\gamma$-finding equations, we assumed the knowledge of the dual variables $\lambda_{\rm P}^*$ and $\lambda_{\rm D}^*$. We will discuss next how to get these dual variables.

\subsubsection{Dual Variables}
\label{Dual_Variables}
The optimum dual variable $\lambda_{\rm P}^*$ must satisfy equation (\ref{Comp_Slackness_Power}). Thus if $\lambda_{\rm P}^*>0$, then we need $S_1^* -P_{\rm{avg}}=0$. $\lambda_{\rm P}^*$ that satisfies the latter equation gives the optimum solution. The same goes for $\lambda_{\rm D}^*$; if $\lambda_{\rm D}^*>0$, then $p_1^* -\frac{1}{\bar{D}_{\rm max}}=0$. These two equations can be solved using a suitable root-finding algorithm (e.g. the bisection method \cite{Numerical_Recipes_Ch9}). To find $(\lambda_{\rm P}^*,\lambda_{\rm D}^*)$, we propose Algorithm \ref{Alg_lambda_P_lambda_D} that executes two nested bisection algorithms. We note that Algorithm \ref{Alg_lambda_P_lambda_D} can be systematically modified to call any other root-finding algorithm (e.g. the secant algorithm \cite{Numerical_Recipes_Ch9} converges faster than the bisection algorithm).

\begin{algorithm}
\caption{Finding $(\lambda_{\rm P}^*,\lambda_{\rm D}^*)$}
\begin{algorithmic}[1]
\label{Alg_lambda_P_lambda_D}
\STATE Initialize $t \coloneqq 1$, $\lambda_{\rm D}^{(1)} \coloneqq \lambda_{\rm D}^{\rm initial}$
\WHILE{$p_1^* -\frac{1}{\bar{D}_{\rm max}} > \epsilon$}
\STATE Initialize $n \coloneqq 1$, $\lambda_{\rm P}^{(1)} \coloneqq \lambda_{\rm P}^{\rm initial}$
\WHILE{$S_1^*-\frac{1}{\bar{D}_{\rm max}} > \epsilon$}
\STATE Update $\lambda_{\rm P}^{(n+1)}$ with the bisection algorithm's update equation \cite{Numerical_Recipes_Ch9}
\STATE $n \leftarrow n+1$
\ENDWHILE
\STATE Update $\lambda_{\rm D}^{(t+1)}$ with the bisection algorithm's update equation \cite{Numerical_Recipes_Ch9}
\STATE $t \leftarrow t+1$
\ENDWHILE
\STATE $(\lambda_{\rm P}^*,\lambda_{\rm D}^*) \coloneqq (\lambda_{\rm P}^{(n)},\lambda_{\rm D}^{(t)})$
\end{algorithmic}
\end{algorithm}


\section{Simulation Results}
\label{Simulation_Results}
\begin{figure}
\centering
{\includegraphics[width=0.95\columnwidth]{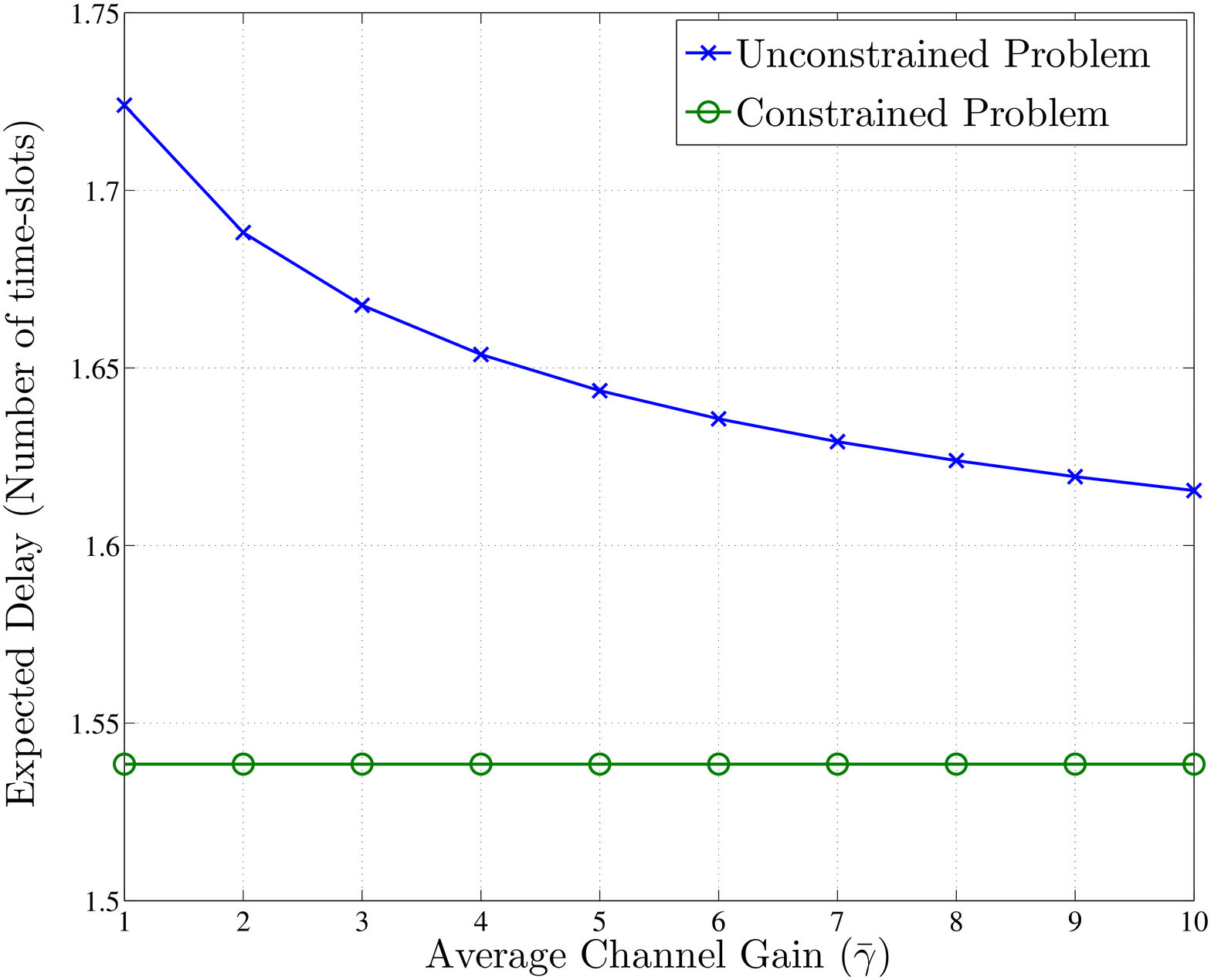}}
\caption{The expected delay is bounded in the constrained maximization problem.}
\label{Delay_Fig}
\end{figure}

\begin{figure}
\centering
 {\includegraphics[width=0.95\columnwidth]{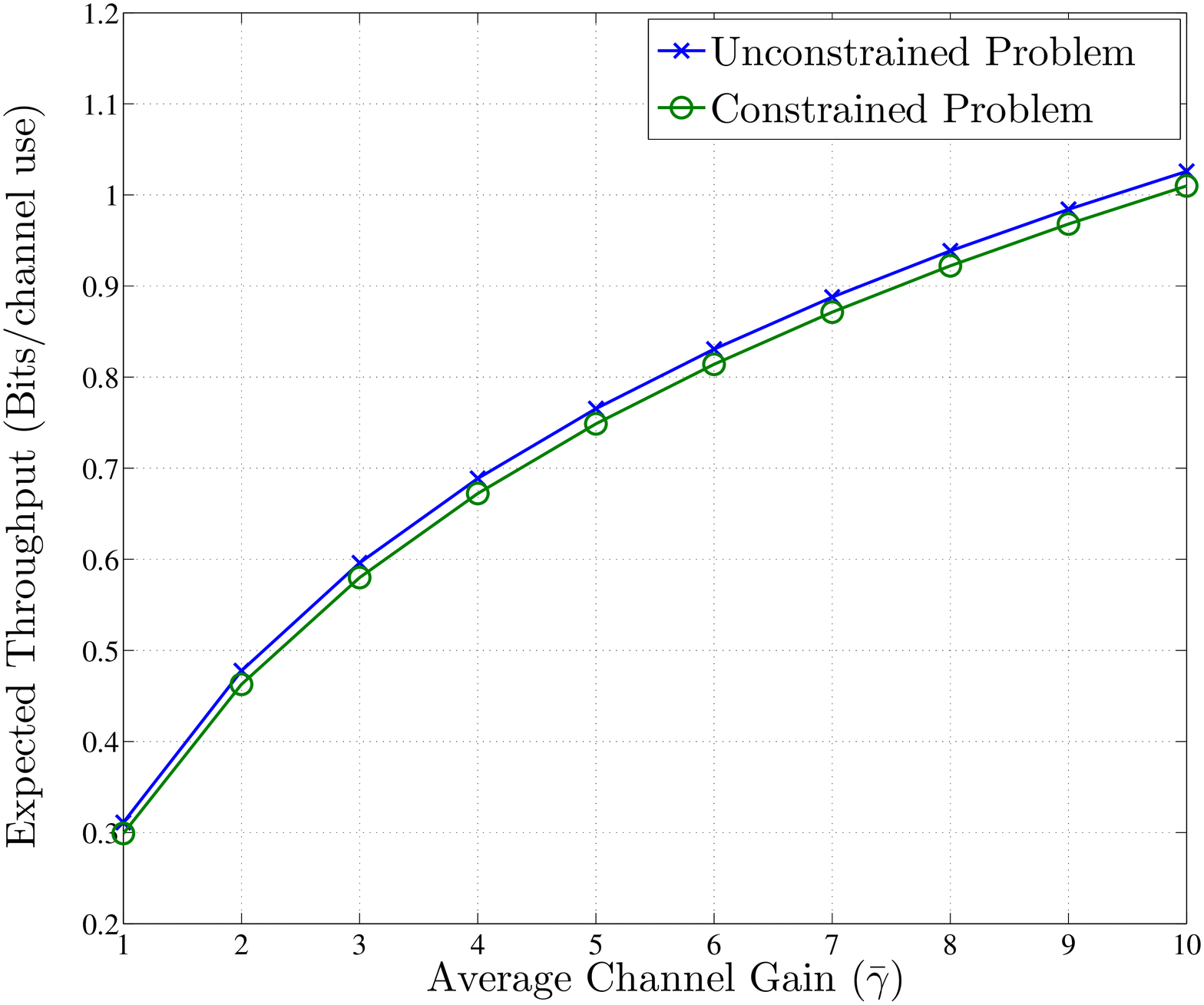}} 
\caption{Throughput of the constrained system compared to the unconstrained one. The proposed model does not sacrifice much throughput. The relative gap difference is less than $4\%$ at $\bar{\gamma} \geq 1$.}
\label{Throughput_Fig}
\end{figure}

The two-level power control system was simulated considering a Rayleigh-fading channel between the secondary transmitter and the intended secondary receiver. The channel has an exponential gain distribution (i.e. $f_\gamma(\gamma)=\frac{1}{\bar{\gamma}}\exp(-\gamma/\bar{\gamma})$, where $\bar{\gamma}$ is the average channel gain of the transmitted assuming unity noise variance). We assumed $M=10$ channels each having $\theta_i=0.1$, $\tau/T=0.05$. To select a suitable value for $\bar{D}_{\rm{max}}$, we note that small values for $\bar{D}_{\rm{max}}$ may lead to an infeasible problem where the system is not able to satisfy this small delay constraint. On the other hand, large values for $\bar{D}_{\rm{max}}$ may yield a trivial solution that can be found by neglecting the delay constraint in our optimization problem. Thus, we set $\bar{D}_{\rm{max}}=1.54T$ which corresponds to the minimum delay that the system can achieve (i.e. $\frac{1}{1-\prod_{i=1}^M\left(1-\theta_i\right)}$). The performance of the system was compared to that of an unconstrained throughput-maximization problem for different values of $\bar{\gamma}$. Fig. \ref{Delay_Fig} compares the expected delay versus the average channel gain $\bar{\gamma}$. In the unconstrained scenario, the expected delay is not controlled and exceeds the maximum average-delay-constraint $\bar{D}_{\rm max}$ that the system can tolerate. On the other hand, when adding the delay constraint to the optimization problem, we guarantee that the delay will be bounded below $\bar{D}_{\rm{max}}$ without sacrificing much throughput. The throughput for both cases is shown in Fig. \ref{Throughput_Fig} where the relative gap is less than $4\%$ for $\bar{\gamma}=1$ and decreases as $\bar{\gamma}$ increases.

\begin{figure}
\centering
{\includegraphics[width=0.95\columnwidth]{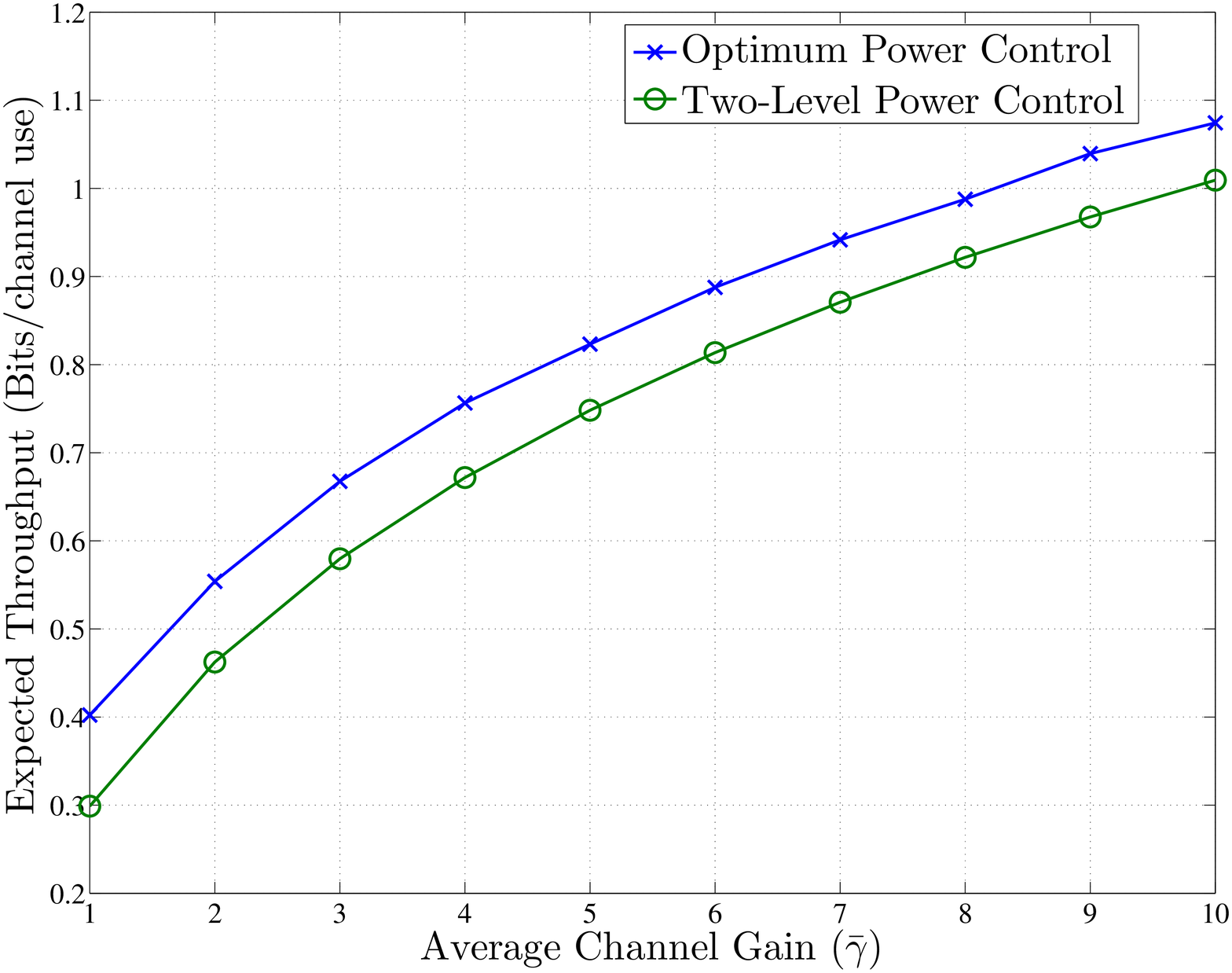}}
\caption{Comparison of the throughput with and without optimal power control. Using optimal power control increases the throughput by more than $6\%$ at $\bar{\gamma} \leq 10$.}
\label{Throughput_Fig_Opt_Pow_Control}
\end{figure}

For the optimum power control scenario, the same channel model was used. For a fair comparison with the two-level power control system, the average power was chosen to be equal to the expected power in the two-level power control system.
Fig. \ref{Throughput_Fig_Opt_Pow_Control} compares the expected throughput of the optimal power control system to that of the two-level power control system. At low average channel gain values the throughput increases by about $34\%$ when allocating the power using the water-filling algorithm. Although the relative gap decreases with the average channel gain, the increase in throughput is still significant (more than $6\%$ at $\bar{\gamma}=10$).

\section{Conclusion and Future Work}
\label{Conclusion}
In this work, we formulate a throughput maximization problem constrained by a bound on the expected delay. We provide an optimal solution that has a closed-form expression for this optimization problem when we adopt the two-level power control strategy. Simulations show that constraining the delay will not degrade the throughput significantly, yet will guarantee average delay bounds to the packets of the CR user. Yet when the optimum power control strategy is adopted, we find an improvement in the throughput of about $34\%$ when the average channel gain is $10$.

In this paper, we neglected the sensing errors of the SU assuming perfect sensing. The problem becomes more interesting when the false-alarm and miss-detection probabilities are considered. This is because these errors result in higher delay since a false-alarm event leads to wasting a potential transmission opportunity, while a miss-detection event results in colliding with the PU's signal leading to the SU's receiver not being able to decode, thus waisting a time-slot.

\bibliographystyle{IEEEtran}
\bibliography{MyLib}

\end{document}